\documentclass{article}

\usepackage{arxiv}

\usepackage[utf8]{inputenc} 
\usepackage[T1]{fontenc}    
\usepackage{hyperref}       
\usepackage{url}            
\usepackage{booktabs}       
\usepackage{amsfonts}       
\usepackage{nicefrac}       
\usepackage{microtype}      
\usepackage{graphicx}
\usepackage{doi}
\hypersetup{hidelinks}
\usepackage[numbers]{natbib}

\title{Diverse features of dust particles and their aggregates inferred from experimental nanoparticles}

\author{\href{https://orcid.org/0000-0002-4182-4052}{\includegraphics[scale=0.06]{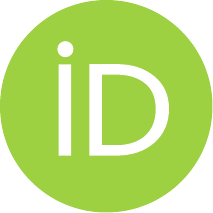}\hspace{1mm}Yuki Nakano} \\
	Institute of Low Temperature Science\\
	Hokkaido University\\
	Sapporo, Japan\\
	\texttt{ynakano@lowtem.hokudai.ac.jp} \\
	\And
	\hspace{1mm}{Yuki Kimura} \\
	Institute of Low Temperature Science\\
	Hokkaido University\\
	Sapporo, Japan\\
    \And
	\hspace{1mm}{Akihiko Hashimoto} \\
	Suns \& Fishermen Society,\\
	Sapporo, Japan\\}

\date{}

\renewcommand{\headeright}{}
\renewcommand{\undertitle}{}
\renewcommand{\shorttitle}{Diverse features of astrophysical dust particles}

\hypersetup{
pdftitle={Diverse features of dust particles and their aggregates inferred from experimental nanoparticles},
pdfauthor={Y. Nakano et al.},
}

\begin{document}
\maketitle

\begin{abstract}
Nanometre- to micrometre-sized solid dust particles play a vital role in star and planet formations. Despite of their importance, however, our understanding of physical and chemical properties of dust particles is still provisional. We have conducted a condensation experiment of the vapour generated from a solid starting material having nearly cosmic proportions in elements. A laser flash heating and subsequent cooling has produced a diverse type of nanoparticles simultaneously. Here we introduce four types of nanoparticles as potential dust particles in space: amorphous silicate nanoparticles (type S); core/mantle nanoparticles with iron or hydrogenised-iron core and amorphous silicate mantle (type IS); silicon oxycarbide nanoparticles and hydrogenised silicon oxycarbide nanoparticles (type SiOC); and carbon nanoparticles (type C), all produced in a single heating-cooling event. Type IS and SiOC nanoparticles are new for potential astrophysical dust. The nanoparticles are aggregated to a wide variety of structures, from compact, fluffy, and networked. A simultaneous formation of nanoparticles, which are diverse in chemistry, shape, and structure, prompts a re-evaluation of astrophysical dust particles.
\end{abstract}




\section{Introduction}
Protoplanetary discs (PPDs) surrounding young stars, are composed of gas and nanometre- to micrometre-sized dust particles, which are derived from molecular clouds at source \citep[reviewed by][]{Williams2011}. While the abundance of dust particles is merely 1 \% in a total mass of PPD medium, they determine the thermal and geometrical structure of the discs and a fraction of them is used for forming planets \citep[e.g.][]{Natta2007, Birnstiel2016}. Studying physical and chemical properties of the particles are crucial to understanding their evolution in PPDs and the planetary systems. 

Interstellar dust is considered a principal source of dust particles before their evolving to solid ingredients in molecular clouds and in PPDs. It is postulated to be mainly composed of either carbon-rich or silicate-rich nanoparticles, depending on their C-rich or O-rich stellar origin either as ejecta of dying-stars or as stellar winds \citep[][]{Tielens2022}. In cold interstellar environment, these nanoparticles are expected to have served as substrates onto which volatile gas adsorbed and evolved into ice or more complex molecules \citep[e.g.][]{Williams1992, Williams2005}. The theoretical picture, interstellar particles have either silicate or carbon cores and volatile, organic and icy mantle \citep[][]{Greenberg1989, Li1997}, is currently in the ascendancy. 

Most interstellar dust, whether it is monomineralic or layered, however, must have been far more processed than was once believed \citep[e.g.][]{Henning2010}. They are destroyed by intense stellar radiation and interstellar shock waves, on a timescale of <10$^8$ yrs \citep[][]{Draine1990, Jones1994, Jones2004}, which is shorter than the lifetime of the galactic interstellar dust ($\sim$5 x 10$^8$ yrs)\citep[][]{Jones2004}. The vaporised refractory atoms and molecules should subsequently be condensed into dust of new generation, viz. recycled, which would refill the interstellar space \citep[][]{Jones2011}. The short timescale for the dust destruction implies that most of interstellar dust possibly lose their primary memory acquired when they formed around stars.

By the same token, dust particle delivered to PPDs can be regenerated one and thus most probably differ from principal particles born around dying stars and in stellar winds. Observations of light scattering by dust particles in PPDs, aided by model calculations, have been applied to understand the properties of the dust particles, which shows that the size of dust particles in PPDs is much larger than that of interstellar dust on average probably due to a dust growth process inside of PPDs \citep[][]{Natta2007}. Fraction of the dust particles might have been reprocessed inside PPDs by vaporisation, melting, mixing, condensation, etc., as inferred from chemical and isotopic evidence in chondritic meteorites \citep[][]{Anders1987} in addition to recycling processes in interstellar environments. The existence of presolar grains in meteorites, however, is a manifestation of survival of original interstellar dust, despite of their abundance as low as a few hundred ppm at most \citep[e.g.][]{Huss1995}.

Apart from the theoretical and observational works, experimental research has also contributed to the study on dust particle properties, especially for interstellar dust \citep[e.g.][]{Stephens1978, Jager1994, Rouille2014, Krasnokutski2017}. Most experimental techniques use simultaneous evaporation and condensation of solid materials in low-pressure ambient gases, and have been applied to various materials such as oxides, silicates, metals, carbon, and metal carbides by way of producing analogues of interstellar nanoparticles and measuring their optical properties \citep[e.g.][]{Jager1994, Jager1998}. The starting materials used in those studies, however, are too simple to represent the composition of actual interstellar and PPD materials. Simple materials are often beneficial for studying elementary kinetics, but there is no guarantee for the nanoparticles thus produced to be real astrophysical analogues. 

We apply a laser flash heating to a starting material which is composed of a homogenised solid charge of the Allende meteorite and of a pure H$_2$ ambient gas. The Allende meteorite contains all natural elements in their cosmic proportions except for the elements which are volatile at room temperature, and is suitable, together with hydrogen atmosphere, as a dust and gas mixture analogous to PPD and interstellar environments. Here we show that diverse kinds of refractory nanoparticles with characteristic chemistry, morphology and aggregation structures condensed simultaneously from the vapour generated by a single shot laser-evaporation and subsequent cooling.

\begin{figure*}
	\includegraphics[width=0.85\textwidth]{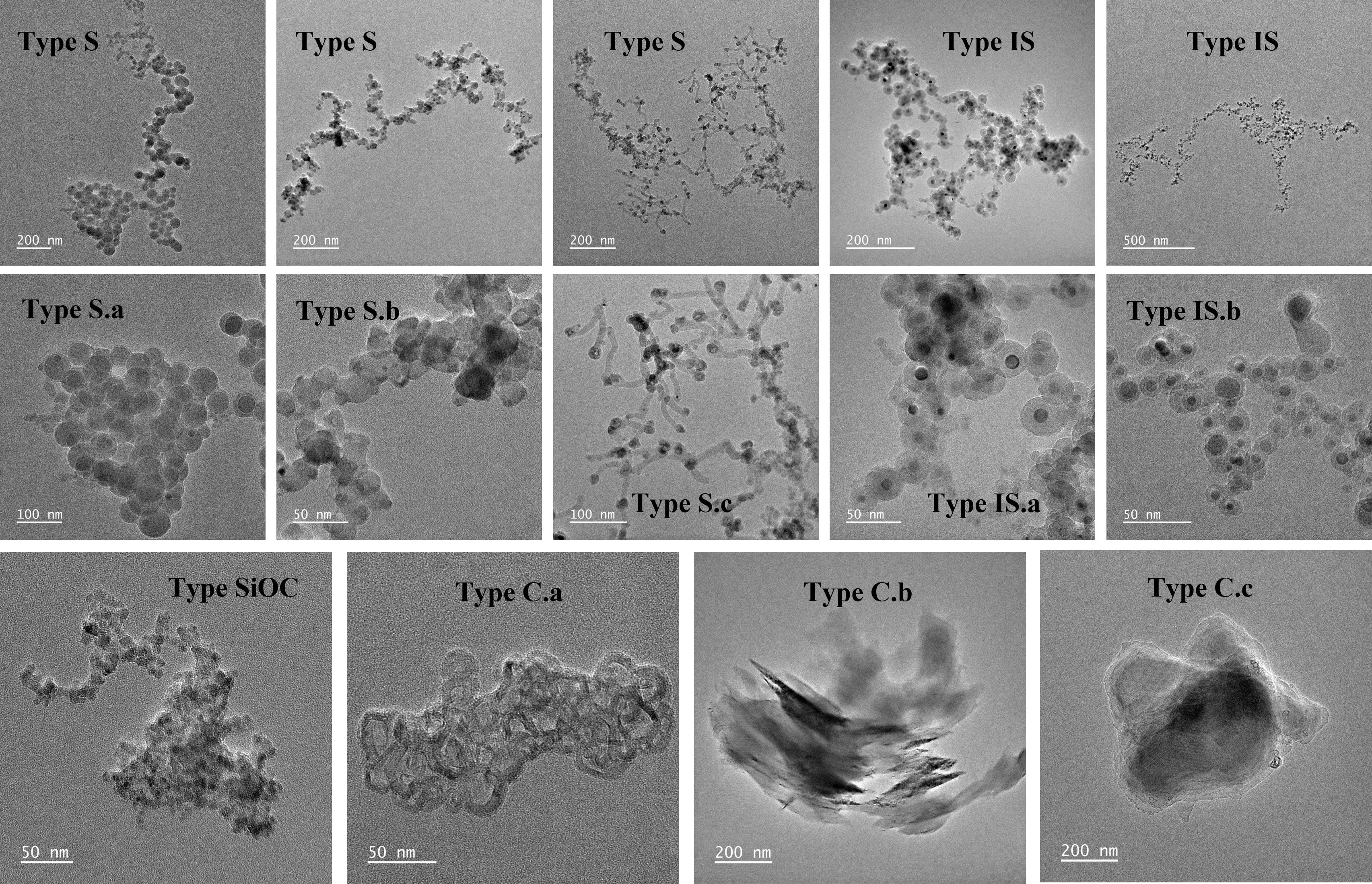}
    \caption{Bright field TEM images of condensed nanoparticle aggregates. Top panels show whole views of aggregates of types S and IS nanoparticles. Centre panels are magnified images of the corresponding top panels. Bottom panels show aggregates of types SiOC and C nanoparticles. Lower case letters following dots indicate sub-types of the preceding NP types. The locations of magnified images in the top panels are shown in Fig. S1.}
    \label{fig:figure1}
\end{figure*}

\section{Experimental and analytical methods}
\label{sec:experiment}
A pulverised and homogenised powder of Allende meteorite (<5 \textmu m in grain size) was used as a source of initial sample. The powder was pressed into cylindrical shape and fired for 12 hours at 900 ℃ in a PO$_2$ controlled furnace with a H$_2$/CO$_2$ = 1/2 gas ratio to stabilise iron as a w\"{u}stite or FeO component in silicates. A narrow hole (0.5 mm in diameter) was drilled to a halfway through the central axis of the sample. Then it was scraped off to its shape, 3 x 2 mm in height and diameter, and weight, 15 mg. A Re wire (0.5 mm in diameter) stuck with a small and thin graphite disc was inserted to the bottom of the hole. The distance between the sample and the disc was kept at $\sim$1.5 mm, so that the sample would neither slip off of the wire nor react with graphite when it got completely molten.

The sample was set in the centre of a vacuum chamber \citep[97 mm inner diameter and 310 mm height; see for details][]{Nakano2020} . A cw-CO$_2$ laser (wavelength 10.6 \textmu m; 20-210 watts; NIIC model NC-30100-M) with a ZnSe lens (focal distance of 127 mm) was used to heat the sample with its hot-spot diameter of 1 mm. The sample was heated with a power of 100 W (temperature $\sim$3000 K) for 3 seconds in a 1 atm H$_2$ gas flow ($\sim$6 litter/min) continuously evacuated with a vacuum pump through a feed-through. A smoke, clusters of condensed particles, rose for a few seconds from the vicinity of the sample. A fraction of the smoke was collected on a standard Cu-grid (3 mm in diameter) set beforehand on the upper inner wall of the chamber, to be served with transmission electron microscopy (TEM) after the run. Nearly ten wt. \% of the original mass evaporated and condensed as nanoparticles (NPs). 

We used a TEM (JEOL JEM-2100F, operated at 200 keV and equipped with an energy-dispersive X-ray spectroscope with a silicon drift detector (JED23000T, JEOL Ltd.)) to characterize morphology, structure, and composition of the collected NPs and their aggregates.

\section{Results}
\subsection{Characterisation of experimental nanoparticles}
\label{sec:result1}
\begin{figure}
    \includegraphics[width=0.33\textwidth]{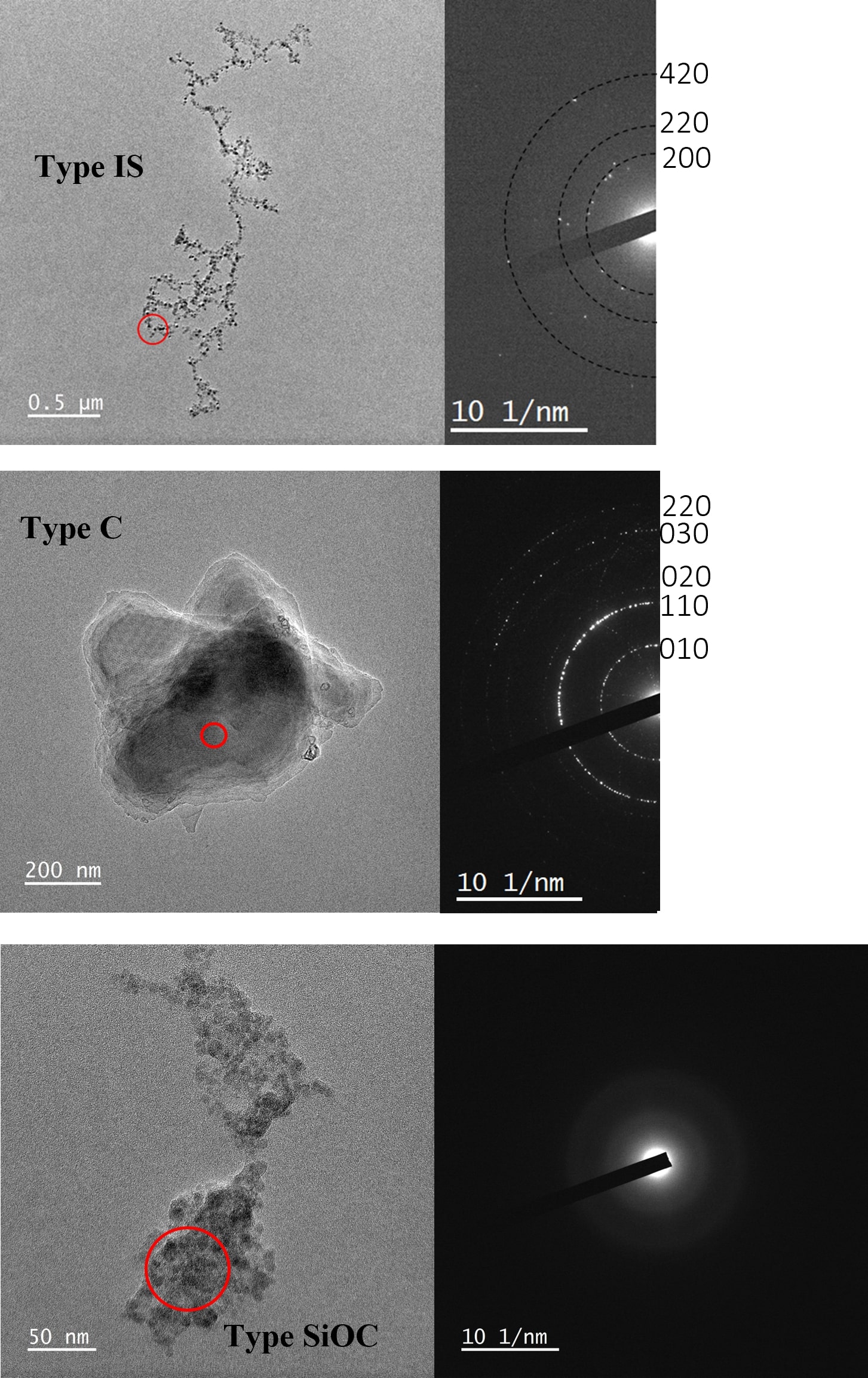}
    \caption{Left panels are bright field TEM images of type IS, C, and SiOC NPs. Right panels show the electron diffraction patterns of the selected areas (marked in red) shown on the left images, indicating $\alpha$-iron \citep[][]{Wilburn1978}, hexagonal-diamond \citep[][]{Bundy1967} and amorphous from top to bottom. Miller indices are also shown for the $\alpha$-iron and the hexagonal-diamond.}
    \label{fig:figure2}
\end{figure}

The nanoparticles (NPs) produced in our experiment invariably exist in their aggregate forms. We found 66 aggregates of NPs on the Cu-grid. On the basis of EDS chemical analysis and electron diffraction, the NPs are categorised into four types. Excluding some cases (see below), each aggregate consists of mono-type NPs. Figures \ref{fig:figure1}, S1, and S2 show aggregates that belong to each of the four types of NPs. Diffraction patterns and typical EDS spectra for each type of NPs are shown in Figs. \ref{fig:figure2} and S3 to S10. Table \ref{tab:table1} summarises physical and chemical properties of the four types of NPs and their aggregates.

\begin{table*}
	\centering
	\caption{Characterisation of the collected condensed nanoparticles.}
	\label{tab:table1}
    \includegraphics[width=0.9\textwidth]{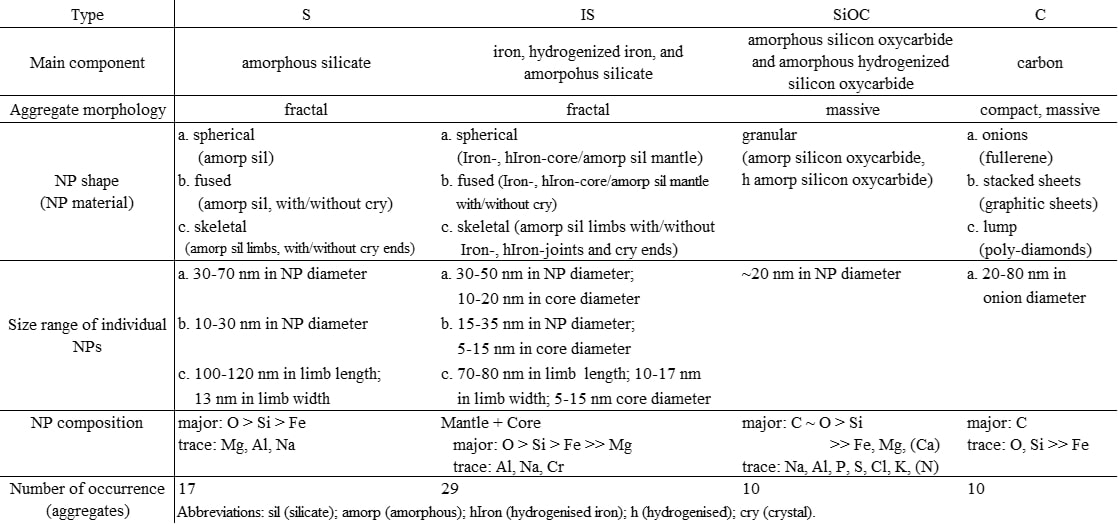}
\end{table*}

Type S are Si-, Fe-rich silicate NPs and are amorphous from their diffraction (Fig. S3) except minor crystals. Type IS are core/mantle NPs with an iron or iron compound core and an amorphous silicate mantle. Silicate are similar in composition to those of type S NPs (Fig. S7). Iron is $\alpha$-Fe (Fig. \ref{fig:figure2}), and the iron compound is most probably a hydrogenised iron from the observation to be described in Section \ref{sec:result2}. For type S and IS NPs we determined the chemical composition of selected NPs by adjusting the size of electron beam diameter to a size of individual NP and thus the obtained composition is a bulk composition of the whole NP grain whether or not it is core-mantle structured. The S and IS types are sub-divided into three subtypes according to their NP morphology: spherical, fused, and skeletal (a, b, and c in Table \ref{tab:table1}, respectively). The b and c subtypes possess nanocrystals such as w\"{u}stite (Fig. S4). Aggregates of either S or IS type NPs tend to be fluffy or networked in their morphology, with many branches of linearly-connected NPs linked together at one or more ends (Figs. \ref{fig:figure1} and \ref{fig:figure2}; see also Fig. S1).

We have found NPs which consistently exhibit Si, O, and C as major peaks in EDS spectra (Fig. S9; we randomly measured many parts of a selected aggregate and took their average composition), and named them SiOC NPs. Diffuse rings in the electron diffraction pattern of type SiOC NPs indictate amorphous strctures (Fig. \ref{fig:figure2}). We identify them as "silicon oxycarbides" from their typical morphology and composition akin to the reported hydrogenised silicon oxycarbides \citep[][]{Tavakoli2015} in addition to the similarity in experimental conditions between ours and \citet{Tavakoli2015}. Silicon oxycarbide refers to a carbon-containing silicate glass wherein oxygen and carbon atoms share bonds with silicon in an amorphous network structure \citep[][]{Pantano1999}. Aggregates of SiOC NPs are rather compact and massive compare to those of type S and IS NPs.

Type C NPs are nearly pure carbon (Fig. S10; we randomly measured many parts of a selected aggregate and took their average composition). They are sub-divided into three subtypes according to their NP morphology and crystal structures; (a) onions (fullerene), (b) stacked sheets (graphite), and (c) lump (poly-diamond) (Figs. \ref{fig:figure1}, \ref{fig:figure2}, S5 and S6; see also Table \ref{tab:table1}). Aggregates of all the subtypes are compact in morphology (Fig. \ref{fig:figure1}).

The sizes range of each type of NPs is listed in Table \ref{tab:table1}. Type S and IS are similar in their size range. It is noted that the size of NPs decreases in the order of subtypes a, b, and c (the width of limbs for the subtypes c) for both type S and IS NPs and that there is a size break between the spherical (subtype a) and the fused (subtype b) at $\sim$30 nm in NP diameter (see Table \ref{tab:table1}). The diameter of SiOC type NPs is fairly uniform $\sim$20 nm. Type C.a NPs are aggregates of onion shaped units whose size is 20-80 nm. Basically the size of all types of NPs is within 10-100 nm (the largest being a limb length of subtype S.c or IS.c NPs) except for type C.b (stacked sheets of graphite) and C.c (poly-diamonds).

There are some exceptions to the mono-NP-type aggregation. Nearly 25 \% of type S and IS aggregates are partially mixed, but such a mixture always occurs in units of separate branches, viz. each branch consists of the same type of NPs. In one instance, an aggregate of type SiOC NPs is attached to a single branch of type IS NPs. Type C NP aggregates are exclusively isolated. These observations suggest that the type S and IS NPs initially formed a two-dimensional, chain-like aggregate of the same NP type and that the chains were linked together later. It must be the latter occasion when mixing of different types of NPs occurred. Type SiOC and C NPs, on the other hand, obviously did not form chain-like structures but agglomerated randomly. 

\subsection{Hydrogenisation of Fe-metal - possibility}
\label{sec:result2}
\begin{figure}
    \includegraphics[width=0.4\textwidth]{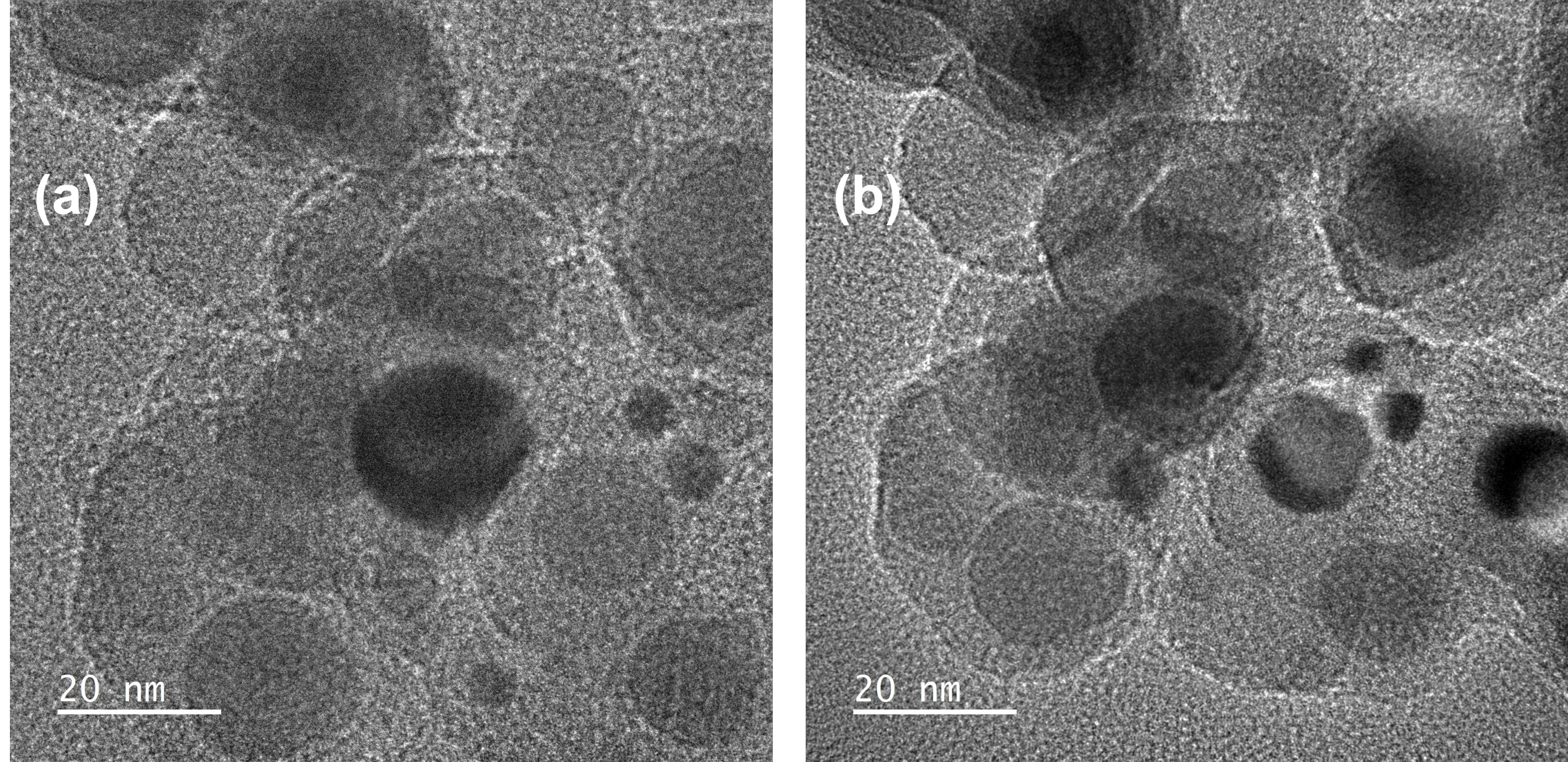}
    \caption{Bright field TEM images of an aggregate of type IS NPs. (a) shows NPs before a prolonged electron beam irradiation for EDS analysis of the area. (b) shows the same NPs after the EDS analysis. Observe that the sizes of both metal cores and silicate mantles of the same NPs diminished after the EDS analysis.}
    \label{fig:figure3}
\end{figure}

During the EDS analysis of type IS NPs, the thickness of both silicate mantle and iron core were significantly reduced after a long exposure to electron beam irradiation while collecting X-ray counts. The diametre of the metal core became smaller by 4 to 20\% (Fig. \ref{fig:figure3}). Such an observation is explained if the core of type IS NPs was hydrogenised beforehand since hydrogen is easy to escape by electron bombardment. The change in core diameter by 4 to 20\% means a reduction of the core volume by 12 to 50\%. According to the experimental works \citep[][]{Carrol1976, Whetten1985, Richtsmeier1985}, Fe atoms and clusters react with hydrogen molecules and form Fe$_m$H$_n$ clusters. Based on a density functional theory, \cite{Takahashi2013} showed that a single gas-phase Fe atom can bind up to ten hydrogen atoms and that a large Fe-H cluster such as Fe$_9$H$_{56}$ is also stable. At present we do not know how many hydrogen atoms per each Fe atom are contained in the type IS NPs. Considering the smallest size of H atoms and the volume shrinkage up to 50\%, however, the number of hydrogen per Fe should be much larger than unity at least in some of type IS NPs. The shrinkage of silicate mantle (also type S NPs) by electron bombardment during the TEM observation also suggests that silicates were highly hydrogenised, perhaps as Si-H or Si-OH bonds.

\section{Discussion}
\subsection{Formation of diverse nanoparticles}
\label{sec:froamtion1}
Our experimental NPs condensed from the vapour of a molten sample in the 1 atm ambient gas of pure H$_2$. According to classical nucleation theory, nucleation of condensed phases from supersaturated vapour proceeds through the formation of gas-phase clusters by mutual collisions of vapour molecules \citep[][]{Castleman1988}. The ambient gas would confine vapour molecules near the evaporation source, here the molten sample. In our experimental condition, the mean free path, approximately 100 nm, could bring a supersaturation of the vapour in the immediate vicinity of a heated solid sample.

Even in a single flash heating event, four types of NPs, viz. S, IS, C, and SiOC were produced in our experiment. According to the classical nucleation theory, each type of gas-phase clusters are produced in a supersaturated vapour around a molten sample (viz. the vapour source), and the clusters could grow through more collisions of the same type of vapour molecules. Then, we would need the reason for the simultaneous formation of many types of NPs. We might consider two possibilities. (1) The condition of the supersaturated vapour such as composition and density is spatio-temporally heterogeneous. (2) In a homogeneous evaporating gas, NPs of the type that become supersaturated early form before other types, and a time-dependent condensation sequence occurs between different types of NPs.

We point out the mode of occurrence of NPs in their aggregates; all aggregates are basically composed of the same type of NPs no matter what the type of NPs is. We have described in Section \ref{sec:result1} that some branches of type IS NP aggregates are occasionally being replaced with branches of type S NP aggregates, but they never mix with each other in individual NP wise. If various types of NPs coexisted nearby in a parcel of the ambient gas, they would collide each other and form aggregates with an assortment of various NPs. In case of the first possibility, a microscopic convection induced by the laser heating would have isolated gas parcels each with unique chemistry, density, and/or temperature, in which mono-type NPs condensed and stuck together. In case of the second possibility, not only the nucleation and growth of one type of nanoparticles precedes those of other types of nanoparticles, but also the mutual sticking of the earliest type nanoparticles must prevent late types of nanoparticles from sticking onto the former. It means that the 
 ticking probabilities between different types of nanoparticles must be significantly small.

\subsection{Formation of type IS nanoparticles}
\label{sec:froamtion2}

Formation of the type IS NPs that contain iron core and amorphous silicate mantle is particularly curious because its iron core seems to form prior to its silicate mantle although a homogeneous nucleation of a metallic iron is highly inefficient \citep[][]{Kimura2017}. In order to nucleate iron first, classical nucleation theory tells, a very high density of iron vapour in the supersaturated vapour and/or a significant rise in iron nucleation rate is required as compared to those of silicates. Such conditions seem unlikely in our experiment. A recently suggested non-classical nucleation process \citep[e.g.][]{DeYoreo2015} has been applied to postulate a nucleation of Ti-C system NP \citep[][]{Kimura2023}. A similar explanation may be conceivable for metal-silicate NPs like ours. Precursory nuclei of silicate may form through homogeneous nucleation and serve as substrate not only for more silicate vapour but also for iron vapour to stick on. At some point, fusion of the numerous nuclei could lead to a single NP that is composed of an iron core covered with a silicate mantle. Such a schematic scenario for the formation of type IS NPs needs a quantitative evaluation.

A hindering factor of Fe metal nucleation is a large binding energy of Fe atoms released during the formation of a dimer, that prompts its dissociation into atoms again \citep[][]{Kimura2017}. If monomers such as FeH$_n$ (n $\geq$1) were available, however, formation of dimers and oligomers would be easier because the binding energy can be released by Fe-H vibrations. Therefore, it is possible that FeH$_n$ forms clusters and (FeH$_n$)$_m$ nucleates. As described in Section \ref{sec:result2}, our experiment, performed in pure 1 atm hydrogen, has a reasonable probability to have formed hydrogenised iron.

In interstellar environment the ambient gas (H$_2$ and H) density is very low. Considering the number ratio of H to Fe in the mass balance between the ambient gas and the vapour, however, our experiment is not so different from the interstellar environment, or even smaller from the following analysis. Since the vapour pressure of the Allende meteorite rock (the staring material) at $\sim$3000 K (the experimental condition) is nearly 0.3 atm \citep[][]{Nakano2020}, the experimental gas/vapour density ratio is $\sim$30 by taking into account the difference in temperature, 300 K (ambient H$_2$) vs. 3000 K (vapour), while the gas/dust mass ratio in interstellar medium is $\sim$100 including ice in the dust. Allende meteorite rock dose not contain ice. Thus, a hydrogenisation of Fe (and also Si) is not unlikely in interstellar medium. In fact, FeH and FeH$_n$ (n>1) molecules are found in solar and other star spectra \citep[e.g.][]{Carrol1976, Wing1977} and in exoplanet atmospheres \citep[][]{Tennyson2012}. In interstellar space and PPD, iron particles could have formed via iron dimers with the help of hydrogen in its initial stage. Which nucleates faster in interstellar space, (hydrogenised) iron or silicate, however, still needs investigation and leaves the possibility of type IS NPs in space on hold.

\subsection{Comparison with interstellar dust}
\label{sec:size}

Interstellar silicate dust imposes broad absorption/emission bands on starlight features at 10 and 18 \textmu m, attributable to the Si-O stretching and O-Si-O bending vibrations \citep[e.g.][]{Henning2010}. The experimental NPs of types S and IS, as well as type SiOC, are candidates for the interstellar silicate dust because not only types S and IS but the SiOC type also has Si-O bonds. For interstellar carbonaceous dust, spectroscopic evidence indicates that the predominant phase is amorphous carbon \citep[reviewed by][]{Ehrenfreund2010}, but type C NPs produced in our experiment show diamond, fullerene, and graphitic structures although all of the allotropic forms of C are found in space \citep[][]{Ehrenfreund2010}. 

The spectral feature of starlight alone is not enough to determine the composition of the interstellar silicate. Minerals found in actual primitive meteorites and a theoretical sequence of mineral condensation from the gas having a cosmic abundance of elements have been considered in order to speculate a more detailed composition \citep[][]{Tielens1998, Tielens2022}. The optical properties of conceivable analogue minerals have been measured in laboratory and compared with the spectra features of observed starlight \citep[e.g.][]{Jager1994}. Our experimental NPs also wait for spectroscopic studies. As \citet{Mathis1989} suggested, the interstellar dust may exist in the form of fluffy aggregates rather than separate nanoparticles. The morphology and optical properties of dust aggregates have been investigated by experimental and modelling works \citep[e.g.][]{Bazell1990, Kozasa1992, Wurm2002}. Since our experimental NPs exhibit a wide variety of aggregate structures, it is imperative to study the effect that such various aggregate structures have on the spectral shapes.

The size distribution of interstellar dust has been estimated by combining the spectral feature of starlight and modelling works \citep[e.g.][]{Kim1995, Siebenmorgen2014}. The size range of amorphous carbon and silicate grains is 0.5 to 500 nm in diameter, depending on the assumption about their shapes \citep[][]{Siebenmorgen2014}. All types of our experimental NPs are within the estimated size range, but their majorities are less than 70 nm (Table \ref{tab:table1}). The shape of the interstellar dust can be evaluated by the polarization of starlight, and it has been predicted that interstellar grain is elongated \citep[e.g.][]{Mathis1990}. Our type S and type IS NPs show both spherical (subtype a) and skeletal (subtype c) shapes and appear to vary continuously from spherical to elongated (Table \ref{tab:table1} and Fig. \ref{fig:figure1}). 

As described, it is likely that the core of type IS NPs are variably hydrogenised. If type IS NPs are major dust in interstellar medium, it may resolve a long-standing paradox that the interstellar medium is highly depleted in gaseous iron and yet solid iron particles are not positively identified \citep[e.g.][]{Savage1979a, Dwek2016}. If the hydrogenised iron had been transferred to the solar system during the formation of planets, their cores would have been rich in hydrogen \citep[][]{Okuchi1997, Yuan2021, Sun2023}. It may also give an easy answer to the origin of terrestrial water, because the D/H ratio of the terrestrial water, $\sim$10$^{-4}$ \citep[][]{Lecuyeri1998}, is close to that of the interstellar gas, $\sim$10$^{-5}$ \citep[][]{Linsky1998}, rather than to that of the interstellar ice, $\sim$2 x 10$^{-2}$ \citep[][fig. 10]{Hashimoto2021}. A simple arithmetic tells that nearly 99 \% of the terrestrial hydrogen would have inherited the interstellar gas if it had been derived only from those gas and ice.

Based on our experimental study that uses a realistic chemical composition, both for dust and gas, of interstellar and PPD medium, we point out the followings. First, the composition of interstellar and PPD dust particles probably has a wide range as shown by the diverse types of nanoparticles generated in our experiment, and are not necessarily stoichiometric such as inferred from meteoritic minerals and from the equilibrium condensation calculation. Second, the shape of nanoparticles and the structure of their aggregates can be variable even if nanoparticles are chemically identical, such as shown by our experimental type S and IS NPs. The spectroscopic features of interstellar and PPD dust and analogue materials to compare with should be evaluated in light of nanoparticles which are much more diverse in chemistry, shape, and structure. Third, it is likely that interstellar and PPD dust particles are "unexpected amalgam" of elements, which otherwise are not easy to combine or coexist, such as the hydrogenised iron core (Fe and H combined) in type IS NPs, the (hydrogenised-) silicon oxycarbide (Si, O, C, and H combined) in type SiOC NPs, and type IS NPs (where metal and silicate intimately coexist, not as a mechanical, chance mixture). Such a feature may partly originate in atomic and molecular interactions in angstrom$\sim$nanometre-scaled clusters and particles, which are different and unpredictable from those in macro-scaled bulk matter. Although the ambient gas was a pure H$_2$ in our experiment, a use of a more realistic (interstellar-like) gas, which also contains He, Ar, and CO and many other volatiles such as H$_2$O, CO$_2$, N$_2$, NH$_3$, H$_2$S, etc. that are mainly ice components at low temperature, will produce more diverse types of NPs by reaction with the vapour generated from a cosmic solid material or Allende meteorite.

\section{Conclusion}
\label{sec:conclusion}
A single laser flash heating in a pure H$_2$ gas of a solid starting material having nearly cosmic proportions in elements generated four types of nanoparticles simultaneously, which are diverse in chemistry, size, shape, and aggregation structure. Mechanisms of simultaneous nucleation of diverse nanoparticles are considered, but micro-scaled fluid dynamics as well as multi-component effects on nucleation and cluster physics need to be more clarified. The incorporation of hydrogen atoms into the iron core in type IS nanoparticles should have a profound consequence on the fate of astrophysical and planetary irons and the origin of planetary water. Spectroscopic studies of nanoparticles and aggregates as we have produced are yet to come in future. Meanwhile, astronomical survey should consider a wider range in chemical and structural diversity of nanoparticles in interstellar environment and in PPDs.

\section*{Acknowledgements}
We acknowledge staffs of Institute of Low Temperature Science and Department of Cosmosciences, Planets \& Space group, Hokkaido University for their technical supports. We have greatly benefited from discussion with Dr. H. Hidaka, Dr. M. Tsuge, and Dr. T. Yamazaki.

\bibliographystyle{unsrt}
\bibliography{references}

\label{lastpage}
\end{document}